\documentclass[reprint,amssymb,amsmath,aps,prb,showpacs]{revtex4-1}
\usepackage[dvips]{graphicx}
\usepackage{bm}
\usepackage{color}
\begin{document}
\title{Spin wave excitations in the pyrovanadate $\alpha$-Cu$_2$V$_2$O$_7$}
\author{A. Banerjee$^1$}
\author{J. Sannigrahi$^1$}
\author{S. Bhowal$^1$}
\author{I. Dasgupta$^1$}
\author{S. Majumdar$^1$}
\author{H. C. Walker$^2$}
\author{A. Bhattacharyya$^{2,3,4}$}
\author{D. T. Adroja$^{2,3}$}
\affiliation{$^{1}$Department of Solid State Physics, Indian Association for the Cultivation of Science, 2A \& B Raja S. C. Mullick Road, Jadavpur, Kolkata 700 032, India }
\affiliation{$^2$ISIS Facility, Rutherford Appleton Laboratory, Didcot, United Kingdom}
\affiliation{$^3$Highly Correlated Matter Research Group, Physics Department, University of Johannesburg, Auckland Park 2006, South Africa}
\affiliation{$^4$Department of Physics, Ramakrishna Mission Vivekananda University, Belur Math, Howrah 711202, West Bengal, India}

\pacs {78.70.Nx, 75.85.+t, 75.30.Ds}

\begin{abstract}
The pyrovanadate $\alpha$-Cu$_2$V$_2$O$_7$  belongs to the  orthorhombic ($Fdd2$) class of crystals with  non-centrosymmetric  crystal structure.  Recently,  the  compound has been identified to be a magnetic  multiferroic   with a substantial  electric polarization below the magnetic transition temperature $T_C$ = 35 K. Here we report the results of our  inelastic neutron scattering (INS) studies  on a polycrystalline sample of  $\alpha$-Cu$_2$V$_2$O$_7$. Our INS data clearly show the existence of dispersive spin wave excitations below $T_C$  with a zone-boundary energy of 11 meV at 5~K. We have analyzed the data using linear spin wave theory, which shows good agreement between the experiment and calculation. The analysis is consistent with the third nearest neighbor exchange interaction playing a dominant role in the magnetism of the material. 

\end{abstract}
\maketitle

\par
The family of pyrovanadates with general formula M$_2$V$_2$O$_7$ (M = Cu, Ni, Co, Mn) has attracted considerable attention due to their fascinating and diverse crystal structures, which are built out of  various extended units of M-O and V-O polyhedra.~\cite{touaiher,zhang1,zhang2,zhang3} They can have  low-dimensional (chain, sheet, honey-comb etc.) to more complex three dimensional structures with  intriguing  magnetic and electronic properties.~\cite{alex,yashima} These pyrovanadates  broadly crystallize in two different groups of structures: {\it thortveitite} (Sc$_2$Si$_2$O$_7$ type) or {\it dichromate} (K$_2$Cr$_2$O$_7$ type).~\cite{zhang3} 

\par
Among the various members of the group, Cu$_2$V$_2$O$_7$ draws special attention both for its structural and electronic aspects. This compound can crystallize in at least four polymorphic phases (namely $\alpha$, $\beta$, $\beta^{\prime}$ and $\gamma$)  with different lattice symmetry (although all thortveitite type).~\cite{alex,calvo,kriv} The $\beta$  phase can be described as a spin 1/2 Heisenberg system on a two dimensional (2D) honeycomb lattice .~\cite{alex} On the other hand, the $\alpha$ phase is better described by two sets of mutually perpendicular zig-zag chains with strong inter-chain magnetic interactions. Here all  Cu$^{2+}$ ions are equivalent with fivefold coordination to oxygen atoms forming a distorted [CuO$_5$] polyhedron. Each distorted polyhedron is linked with another two via edge sharing and  together they form the two sets of mutually perpendicular zig zag chains (see Fig. 1) in the $bc$ plane.~\cite{calvo,sanchez} Notably, $\alpha$-Cu$_2$V$_2$O$_7$ crystallizes in the Orthorhombic $Fdd2$ structure and it  is the only member of the pyrovanadates to have a non-centrosymmetric crystal structure. 

\begin{figure}[t]
\centering
\includegraphics[width = 9 cm]{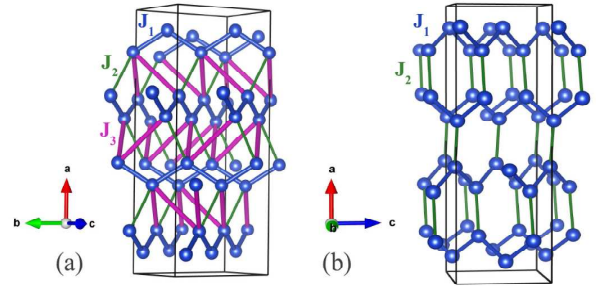}
\vskip 0.4 cm
\caption {(color online)  (a) Various spin exchange interactions in $\alpha$-Cu$_2$V$_2$O$_7$  in a conventional unit cell according to model-I. (b)  The helical honeycomb network (model-II) of spin exchange comprising $J_1$ and $J_2$ only.}
\end{figure}

\begin{table}[t]

	\centering
			\begin{tabular}{|c||c|c||c|}
\hline
 $J$'s in meV & Model-I & \multicolumn{2}{c|}{Model-II}\\
 \cline {3-4}
  & &$J_1 > J_2$&$J_2 > J_1$\\
\hline\hline
$J_1$ & -4.67  & -5.79  &-4.10\\
$J_2$ & 0.8 & -2.61 &-6.31\\
$J_3$ & -9 & -- &--\\
\hline

\end{tabular}

\caption{Various Cu-Cu exchange parameters used for the spin wave simulations. Model-I is based on the  exchange interactions obtained by DFT calculations~\cite{sannigrahi} with required modification, whereas model-II represents the optimized values of $J$'s as obtained from QMC simulations of the bulk magnetization data.~\cite{thai}} 
\end{table}

\par
The magnetic and electric properties of $\alpha$-Cu$_2$V$_2$O$_7$ are equally interesting. The compound undergoes long range magnetic ordering below $T_C$ = 35 K with a spin canted structure.~\cite{sanchez,ponomarenko,Pommer} Recent work by our group  identified the compound  to be an improper multiferroic with the simultaneous development of spontaneous electric polarization and magnetization below $T_C$.~\cite{sannigrahi} Density functional  theory (DFT) based  calculations indicate that the  magnetism in $\alpha$-Cu$_2$V$_2$O$_7$ is a consequence of ferro-orbital ordering due to the unique pyramidal CuO$_5$ environment and  the origin of the giant ferroelectric polarization is primarily due to the symmetric exchange-striction mechanism. In the DFT calculation, the dominant interaction is obtained between the third nearest neighbors of Cu-ions ($J_3$ = -13.61 meV) linking the two mutually perpendicular chains. The other sizable interactions are first nearest neighbor Cu-Cu linkage ($J_1$ = -4.67 meV) along the zig-zag chain via oxygen and the second nearest neighbor inter-chain interaction ($J_2$ = +4.07 meV) along the $a$ axis (see Fig. 1 (a)). Evidently, $J_1$ and $J_3$ are antiferromagnetic (AFM) in nature while $J_2$ is ferromagnetic (FM). Model-I, detailed in Table-I, is based on this scenario with required  modifications of the exchange parameters to suit our Monte Carlo simulations. On the other hand, a quantum Monte Carlo (QMC) based calculation finds a better fit to the experimental magnetization data considering $J_1$ and $J_2$ (both AFM in nature) to be the only  dominant spin-spin interaction terms, which constitute a helical honeycomb-like  2D spin network (Fig. 1(b)) in the $ac$ plane.~\cite{thai} It is to be noted that in the former scenario (where $J_3$ is dominant), the system is magnetically three dimensional, while in the later case (as described by model-II in Table-I) it can be described as a quasi-2D system  with anisotropic magnetic interactions in the honeycomb plane ($J_1 \neq J_2$). The spin canting in $\alpha$-Cu$_2$V$_2$O$_7$ is found to be related to the prevailing strong Dzyaloshinski-Moriya (DM) interaction as described by various authors.~\cite{sannigrahi,thai,epl}
\par 
Inelastic neutron  scattering (INS) is an important tool to study the low-lying excitations in quantum spin systems, where the dispersion
of the excitations and their intensity can provide important information regarding the possible exchange interactions and their relative strengths. In this paper we present INS  results from polycrystalline samples of $\alpha$-Cu$_2$V$_2$O$_7$.  The experimental result is corroborated by spin wave simulations based on classical Monte Carlo and linear spin wave theory, which provide support for the dominance of the third nearest neighbor interaction in the system.

\begin{figure}[t]
\centering
\includegraphics[width = 7 cm]{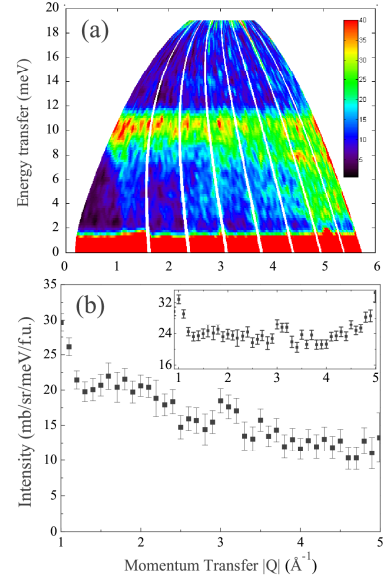}
\caption {(color online) (a) 2D map of the scattering intensity of $\alpha$-Cu$_2$V$_2$O$_7$ as a function of energy transfer ($\hbar\omega$) and momentum transfer ($|{\vec Q}|$) obtained at 5 K. The color scale shows the scattering intensity ($S(|{\vec Q}|$, $\omega$) in absolute units, mb/sr/meV/f.u. (b) shows the scattering intensity  as a function of momentum transfer with phonon scattering corrections applied to the data with the intensity being  summed over the energy range 8 - 12 meV at 5 K. The inset in (b) shows the same data without phonon correction.}
\end{figure}
\par
The details of the preparation of polycrystalline samples of $\alpha$-Cu$_2$V$_2$O$_7$ can be found elsewhere.~\cite{sanchez} The sample was characterized by powder x-ray diffraction and magnetization measurements. INS  measurements were carried out at the  MARI time of flight chopper spectrometer~\cite{mari} at the ISIS pulsed neutron facility, Rutherford Appleton Laboratory, UK.  The powder sample of $\alpha$-Cu$_2$V$_2$O$_7$ was placed in a aluminum foil packet in the form of annulus of diameter 40 mm and height 40 mm and sealed in a thin aluminum can, which was cooled in He-exchange gas using a He closed cycle refrigerator down to a base temperature of 4.5 K. The INS spectra were obtained using different incident energies ($E_i$) between 8 meV and 150 meV as well as different temperatures between 5 K and 100 K.  In order to obtain the scattering intensity in the units of cross section, mb/sr/meV/f.u.,  Vanadium spectra were recorded for the same $E_i$ values. The  simulation of the spin wave excitations were performed using the SpinW software package.~\cite{spinw}

\par
Fig. 2(a) shows the  2D color plot of the neutron scattering data at 5 K with an incident energy of $E_i$ = 20 meV. The elastic line contains Bragg peaks, which can all be indexed on the basis of the crystal structure of $\alpha$-Cu$_2$V$_2$O$_7$. Due to the polycrystalline nature of the samples, the scattering function $S(|{\vec Q}|$, $\omega$) is the powder average of the spin-spin correlation function $S({\vec Q}$, $\omega$), and it does not carry the information regarding the direction of ${\vec Q}$. A band of scattering intensity is observed around the energy transfer  $\hbar\omega$ = 10~meV. The magnetic character of the scattering is evident from the decreasing intensity with increasing $|{\vec Q}|$ in the low ${|\vec Q|}$ region ($\leq$ 3 \AA$^{-1}$). However, the band of intensity is still present at high ${\vec Q}$ ($\geq$ 3 \AA$^{-1}$) values, where one would expect vanishing $S(|{\vec Q}|$, $\omega$) had it been purely magnetic in origin. This is quite clearly visible from the inset of Fig. 2(b), where the raw $S(|{\vec Q}|$, $\omega$) is depicted as a function of momentum transfer for the energy window 8 - 12 meV. We presumed that the high $|{\vec Q}|$ intensity  is purely phononic in origin and calculated the vibrational part of scattering at 5 K using  the 300 K data (well above $T_C$ of the sample) after proper normalization with the Bose factor: $$\mathcal{B}(\hbar\omega) = \frac{1}{1-\exp(-\hbar\omega/k_BT)}$$ The vibrational part is subtracted  from the measured data and the ${|\vec Q|}$ dependence of the scattering intensity is plotted in the main panel of Fig. 2 (b). The resulting data clearly show diminishing intensity with increasing $|\vec {Q}|$ and establishes the magnetic character of the excitation.

\begin{figure}[t]
\centering
\includegraphics[width = 9 cm]{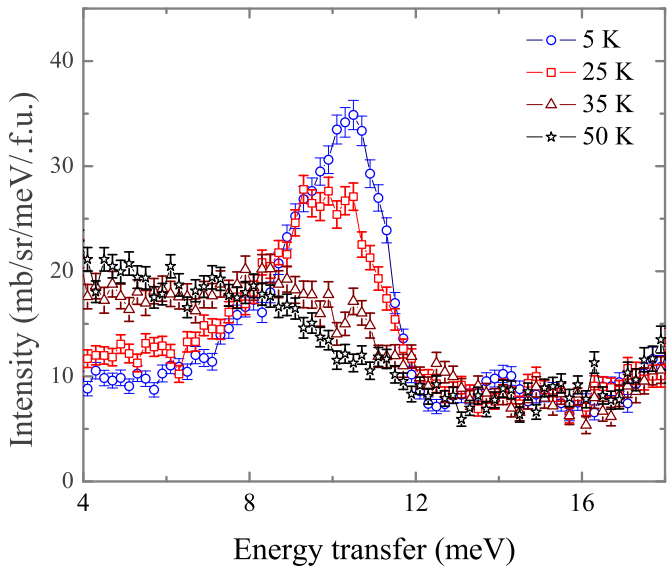}
\caption {(color online) (a) Scattering intensity (without phonon correction) as a function of energy transfer for the window of momentum transfer 0 - 3 \AA$^{-1}$ recorded at different temperatures.}
\end{figure}

\begin{figure}[t]
\centering
\includegraphics[width = 9 cm]{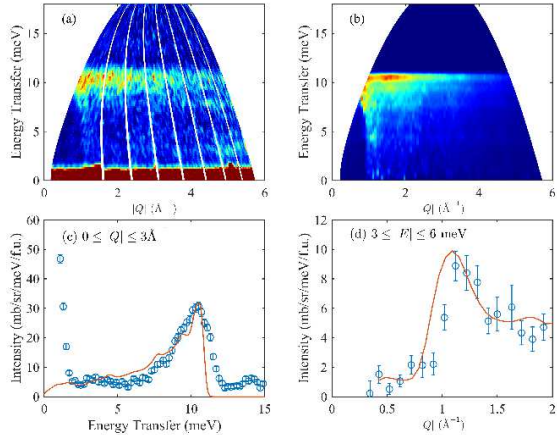}
\caption {(color online) (a) and (b) show the 2D color plots of the neutron scattering intensity as obtained from phonon corrected measured experimental data (at 5 K) and simulated by the SpinW package respectively. (c) and (d) show the experimental and simulated scattering intensity as a function of energy transfer for the $|\vec{Q}|$ range 0-3 \AA$^{-1}$, and as a function of momentum transfer summed over 3-6 meV respectively.}
\end{figure}

\par
We  also investigated the temperature variation of the scattering intensity particularly focusing the low-$|{\vec {Q}}|$ part (0 - 3 \AA$^{-1}$) as depicted in Fig. 3. The peak around 10 meV diminishes with increasing temperature and almost disappears above about $T_C$ = 35 K. This implies that the low-$|{\vec {Q}}|$ scattering  is related to the long range magnetic ordering of the material and that it arises from the spin wave excitations in the ordered state of the material. A small but nonzero scattering intensity is present even in the 50 K data, which is likely to be due to the short range magnetic correlations in the system. This is in agreement  with the  previous magnetization studies where the  deviation from Curie-Weiss law was observed from below about 80 K,~\cite{sannigrahi} indicating precursor local spin-spin correlations to the long  range magnetic order. It is to be noted that there are also phonon modes near the  spin wave energy. 
 
\par
In order to model the experimentally observed magnetic scattering, we calculated the spin wave dispersion and corresponding neutron scattering intensity using the SpinW software package. $\alpha$-Cu$_2$V$_2$O$_7$  contains  magnetic Cu$^{2+}$ (3$d^9$, $S$ = $\frac{1}{2}$) and nonmagnetic V$^{5+}$ (3$d^0$, $S$ = 0) metal ions and therefore only the interactions between Cu$^{2+}$ ions need to be considered. The spin Hamiltonian can be considered as
\begin{equation}
 \mathcal{H} = -\sum_{ij} J_{ij} (\vec{S_i} \cdot \vec{S_j}) + \sum_{ij} \vec{D_{ij}} \cdot (\vec{S_i} \times \vec{S_j}),
  \end{equation}
where the first part represents the symmetric Heisenberg interaction while the second antisymmetric part is related to the DM interaction. For simplifying the calculation, we have not considered any anisotropy term in the Hamiltonian. For the simulation of the spin wave spectra we  used  the canted magnetic structure reported by Lee {\it et al.},~\cite{epl} and exchange  parameters based on model-I~\cite{sannigrahi} and model-II~\cite{thai} as given in table-I. For model-I, as discussed before, the dominant term is $J_3$, while in model-II, the exchange paths comprise a helical honeycomb network with $J_1$ and $J_2$ being the only interaction terms. In the case of model-II, two scenarios may arise, namely $J_1 > J_2$ and $J_1 < J_2$.

\par
At the beginning, we only considered the symmetric exchange terms (namely $J$'s) and ignored the DM terms ($D$'s). In Fig. 4 (a), we have shown the 2D color plot of the experimental INS data where the phonon contribution has been removed. It is clearly seen that the scattering around 10 meV is dominant at low-$|{\vec Q}|$, which is due to its magnetic origin. The simulated scattering data using the $J$ values from model-I (see table-I)  is depicted in Fig. 4 (b), and an excellent agreement is obtained when compared with the observed scattering. This is better viewed via the intensity versus energy transfer  and intensity versus momentum transfer plots (Figs. 4(c) and 4(d) respectively). The experimentally determined value of $J_3$, to match the zone-boundary energy, is less than that found using DFT~\cite{sannigrahi}, but the signs of all three interactions are preserved, and $J_3$ remains dominant. The $J_3$ interaction between two Cu atoms is through two oxygen O atoms (Cu-O-O-Cu) belonging to VO$_4$ tetrahedra. Although V is nonmagnetic, it can expedite the magnetic interaction by electron transfer via its empty 3$d$ level.

\par
The presence of the DM  interaction in $\alpha$-Cu$_2$V$_2$O$_7$ has already been predicted and it is in line with the observed weak ferromagnetism below $T_C$. We  also performed our simulation considering the DM interaction terms  within the framework of  model-I~\cite{sannigrahi}. The introduction of the DM terms does not improve the overall agreement with the experimental data.  In order to investigate the strength of the DM interaction and its role in the magnetic scattering, spin wave measurements on a single crystal of $\alpha$-Cu$_2$V$_2$O$_7$ are highly desirable.
\begin{figure}[t]
\centering
\includegraphics[width = 9 cm]{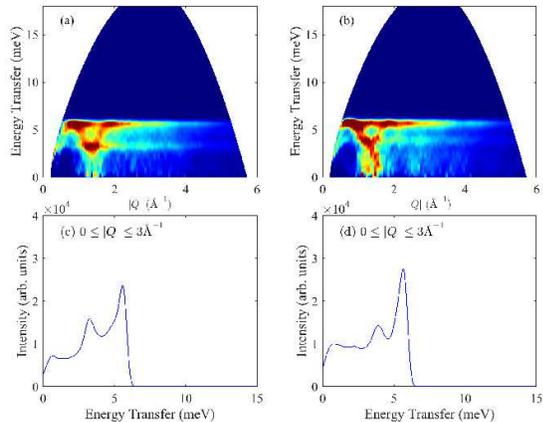}
\caption {(color online) Simulated 2D color plot   for neutron scattering intensity ((a) and (b)) as well as its energy variation ((c) and (d)) considering the Heisenberg interaction terms $J_1$ and $J_2$ only (model-II). For (a) and (c) we have considered $J_1 > J_2$ while for (b) and (d) it is $J_1 < J_2$, using the values quoted in table-I.}
\end{figure}
\par
Despite the generally good agreement between the model-I simulation and the experimental data, the line-width of the cut through the data is found to be wider than the instrumental resolution, which was used for the simulation (Fig. 4(c)). There are several reasons for line-width broadening, such as (i) magnon-electron (ii) magnon-magnon or (iii) magnon-phonon interactions,~\cite{manuel} which are not considered in our simulation. $\alpha$-Cu$_2$V$_2$O$_7$ being a very good insulator, magnon-electron scattering can be neglected. The role of magnon-magnon scattering in line-width broadening can be ruled out as it is only relevant at high temperature close to $T_C$. Being a magnetic multiferroic, the likely origin of the line-width broadening is  magnon-phonon scattering. In multiferroics where  both time and space inversion symmetries are absent, the coupling between low lying magnetic and lattice excitations is relevant.~\cite{rva,sushkov} This is also consistent with the exchange-striction mechanism proposed for the system.~\cite{sannigrahi} The presence of magnon-phonon scattering is also evident from the raw 2D color plot (Fig. 2 (a)), where a high $|\vec{Q}|$ phonon excitation is visible at the same energy as that of the spin wave peak. 
 
\par
We  next considered SpinW simulations for model-II, where $J_1$ and $J_2$ are the only significant exchange terms. The simulated data with two different scenarios ($J_1 > J_2$ and $J_1 < J_2$) are depicted in Fig. 5. Clearly the agreement between the experimental and the calculated data is poor here. The magnetic scattering intensity peaks  around 5 meV  as compared to 10 meV in the observed data. 

\par
The present INS study and subsequent simulations on  $\alpha$-Cu$_2$V$_2$O$_7$ indicate that the analysis is compatible with the model-I, having dominant third nearest neighbor exchange interaction $J_3$. The presence of strong $J_3$, in addition to $J_1$ and $J_2$, makes the spin model three dimensional, resulting in a long range  magnetic ordering below $T_C$.  The multiferrocity in the $\alpha$ phase is induced by the exchange striction and is therefore directly related to this long range ordering.           

\par
In conclusion, we have investigated the  pyrovanadate compound $\alpha$-Cu$_2$V$_2$O$_7$ using inelastic neutron scattering, along with a spin wave analysis. Our INS study reveals well defined dispersive spin wave excitations with a zone boundary energy of 11 meV at 5 K. The spin wave excitations' energy renormalizes with increasing temperature, but the excitations can still be seen at 50 K, which is well above $T_C$, suggesting the presence of short range magnetic correlations. Our spin wave analysis has given a reasonably good description of the experimental data. Furthermore our estimated values of the exchange parameters suggest that the third nearest neighbor Cu-Cu interaction is the dominant one, in agreement with those calculated theoretically using DFT(see Ref.~\onlinecite{sannigrahi} and references therein). We observe some line-width broadening of the spin wave excitation data which can be attributed to the magnon-phonon scattering, a mechanism that is a prevailing factor for mutliferroicity and magneto-dielectric properties in insulating oxides. The present study can foster research on the magnetic excitations in this class of pyrovanadates, and would generate theoretical interest of the development of a more realistic model to understand the complex magnetic and multiferroic behavior of these materials.

\par
The technical help of Dr. Ross Stewart on MARI spectrometer during the experiment is gratefully acknowledged.  A. Bhattacharyya would like to thank DST India for support under Inspire Faculty scheme, while A. Banerjee acknowledges DST-INSPIRE     for Ph.D. research program. 
S. Bhowal thanks Council of Scientific and Industrial Research (CSIR), India for her research support.

\end{document}